\documentclass[12pt]{article}
\usepackage[cp1251]{inputenc}
\usepackage{epsfig}

\begin{document}

\begin{center}
{\large\bf Linear dynamics of charged particles in the main lattices}
{\large of storage rings}

\medskip

 O.E. Shishanin\footnote{shisha-n@msiu.ru}
\medskip

Moscow State Industrial University, Moscow 115280, Russia
\end{center}
\medskip

key words: quadrupole, magnetic field, cell, differential equation, asymptotics, betatron oscillations


\bigskip

\begin{center}
ABSTRACT
\end{center}

 To  study the characteristics of synchrotron radiation in magnetic
 fields of accelerators first the author was necessary to obtain a continuous
 solutions of  Hill's equation. For this purpose the
 gradient or the components of magnetic field were developed in a
 series. The same procedure is followed now in the case of storage
 rings. This approach proved to be interesting not only from the point 
 of view of describing the motion of particles in ordinary three-dimensional space
 but also in the fact that we get new differential equations. 
This brief review can be ragarded as an introduction to the proposed approach.
The next step may be to add nonlinearities. This would be the best approximation to
the determination of betatron oscillations in the existing accelerators.

\newpage

In the beginning, we want to
 note that the asymptotics found in this manner  helped to reveal the experimentally
observed dependence for the spectral and angular formulas of radiation on 
the vertical betatron oscillations [1,2].

   There are three main types of
  lattice for storage rings [3,4]. Among them are the FODO, the
  Chasman-Green lattice(DBA), and triplet achromat(TBA). The FODO
  lattice is the most used. Here one cell is formed of
  separated focusing(F) and defocusing(D) quadrupoles, bending
  magnets(O). Let quadrupoles have a length $a$ and magnets are length $d$.
Then path  of single period is $L=2a+2d+4l$, where $l$ is the
 extension of free shifts. The closed trajectory has scale
$S=2\pi R+(2a+4l)N$, where $R$ is the radius of dipole magnets
and $N$ is the number of periods. Let  average radius is defined
as
$$
2\pi R_0=2\pi R+2(a+2l)N,
$$
where $2\pi R=2dN$. This can be written as $R_0=(1+k)R$, where
  $k=(a+2l)/d$.

The magnetic field of dipole is $B_z=B$ and components of
quadrupole define as
$$
H_z^f=-gx, H_z^d=gx,  H_x^f=-gz, H_x^d=gz,
$$
where $g$ is the lens constant, index $f$ and $d$ means focusing
and defocusing. $x$ coordinate  coincides with the radial direction.

The devices of one period give approximately for an axial part of magnetic field
the following alternation:
$$
-gx, \quad \varphi\in[0, aT];
$$
$$
B, \quad \varphi\in [(a+l)T, (a+l+d)T];
$$
$$
gx, \quad \varphi\in [\pi/N, (2a+2l+d)T];
$$
$$
B,  \varphi\in [(2a+3l+d)T, (2a+3l+2d)T],
$$
where $T=2\pi/(NL)$, $\varphi$ is the azimuth angle.

The  Fourier series expansion of this magnetic fields takes the
form
$$
H_z=\frac{2d}{L}B+\sum_{k=1}^{\infty}[\frac{4B}{\pi}
\frac{(-1)^k}{k} \cos\frac{\pi k}{2}\sin\frac{\pi k}{L}d-
$$
$$
\frac{2}{\pi}gx\frac{1-(-1)^k}{k}\sin k\tau_1] \cos
k(\tau-\tau_1),
$$
where $\tau=N\varphi, \tau_1=\pi a/L$. In particular, in the study
of a radiation problem the dipole magnetic field can be averaged
and then
$$
\bar H_z=\frac{2d}{L}B-gxf(\tau),
$$
where $ f(\tau)=(4/\pi) n(\tau)$,
$$
 n(\tau)=\sum_{\nu=0}^\infty{f_{2\nu+1}}
\cos(2\nu+1)(\tau-\tau_1),\qquad
f_{2\nu+1}=\frac{\sin(2\nu+1)\tau_1}{2\nu+1}.
$$
If $n(\tau)$ is differentiated a divergent series is arised. The
second component of field is  derived as $H_r=-gz f(\tau)$.

Based on previous observations, we determine
the angular velocity  in the following form:
$$
\dot
\varphi=\frac{\omega_0}{1+k}(1-\frac{x}{R_0}+\frac{3}{2}\frac{x^2}
{R_0^2})+ \eqno (1)
$$
$$
\frac{\omega_q}{R_0^2}\int f(\tau)(z\dot z-x\dot x)dt,
$$
where
$$
\omega_0=\frac{e_0B}{m_0c},\quad \omega_q=\frac{e_0 gR_0}{m_0
c},\quad 2d/L=1/(1+k).
$$

In linear approximation the equations of betatron oscillations
become
$$
\frac{d^2z}{d\tau^2}+\frac{1}{N^2}\frac{(1+k)\omega_q}
{\omega_0}f(\tau) z=0, \eqno (2)
$$
$$
\frac{d^2x}{d\tau^2}+\frac{1}{N^2}[1-\frac{(1+k)\omega_q}
{\omega_0}f(\tau)] x=0. \eqno (3)
$$

 Let us introduce new constants
$$
C_1=\frac{gR_0(1+k)}{B}, \qquad \lambda^2=\frac{4C_1}{\pi N^2}.
$$

Under the existing $g$ and $B$ it turns out that parameter
$\lambda>>1$. Eq. (2) can be rewritten as
$$
\frac{d^2z}{d\tau^2}+\lambda^2 n(\tau) z=0. \eqno (4)
$$
 Expression (4) is the Hill equation with a large parameter.
 It is also differential equation with periodic coefficient and
with a small parameter at the highest derivative. By way of illustration several
procedures of solution for Eq.(4) will be tested.

   To find the first solution we can  take
 the WKB-method [5]. Let us  put
$$
z=exp(i\lambda G(\lambda,\tau))\cdot \varphi(\tau),
$$
where $G(\lambda,\tau)=G_0(\tau)+G_1(\tau)/\lambda+...$
 In this case instead  of Eq.(4) we have
$$
\frac{d^2\varphi}{d\tau^2}+2i\lambda\frac{dG}{d\tau}\frac{d\varphi}{d\tau}+[i\lambda
\frac{d^2G}{d\tau^2}-\lambda^2(\frac{dG}{d\tau})^2+\lambda^2n(\tau)]\varphi=0.
$$
Bearing in mind here terms with $\lambda^2$, we obtain
$G_0=\int\sqrt{n(\tau)} d\tau$. The following approximation gives
$\varphi=1/\sqrt[4]{n(\tau)}$. Thus, solution is
$$
z=(1/\sqrt[4]{n(\tau)})\cdot\exp(i\lambda\int\sqrt{n(\tau)}d\tau)+...
$$
But the solution may only depend on the double integral $\int\!\!\int
n(\tau)d\tau d\tau$.

According to [5] again let us use substitution $u=\lambda^2 \tau$
which  changes Eq.(4) as
$$
\frac{d^2z}{du^2}+\frac{1}{\lambda^2}\cdot
n(\frac{u}{\lambda^2})z=0.
$$
It  may be  formally solved as an equation  with a small parameter.
Assume that the solution has the form
$$
z=\exp(i\gamma\, u)\cdot \varphi(u),
$$
where
$$
\varphi(u)=\varphi_0+\frac{1}{\lambda^2}\varphi_1+\frac{1}{\lambda^4}\varphi_2+...,
\quad
\gamma=\frac{\gamma_1}{\lambda^2}+\frac{\gamma_2}{\lambda^4}+...
$$
Note that in the region of stability the value of $\gamma$ must be real.
At various powers of $\lambda$ we shall obtain $\varphi_0=C$,
where $C$ is the constant, $\varphi_1=C\lambda^4C_{21}$, where
$$
C_{nk}=\sum_{\nu=0}^{\infty}\frac{f_{2\nu+1}}{(2\nu+1)^n}\cos
k(2\nu+1)\tau_c
$$
with $\tau_c=u/\lambda^2-\tau_1$. Besides, eliminating the secular
terms we find
$$
\gamma_1^2=\frac{1}{2}\lambda^4\sum_{\nu=0}^\infty\frac{f_{2\nu+1}^2}{(2\nu+1)^2}
\quad \mbox{and}\quad \gamma\approx \frac{\pi^2
a}{4L}\sqrt{1-\frac{4a}{3L}}.
$$
Asymptotics  with an additional term  $\varphi_2$ takes the form
$$
z=C\exp(i\frac{\gamma_1}{\lambda^2}u)[1+\lambda^2C_{21}-2i\gamma_1\lambda^2S_3+\frac{1}{8}
\lambda^4C_{42}+\frac{1}{8}\lambda^4 S_{\mu\nu}], \eqno (5)
$$
where
$$
S_{\mu\nu}=\sum_{\mu=0}^\infty\frac{f_{2\mu+1}}{(2\mu+1)^2}\sum_{\nu=0}^\infty
f_{2\nu+1}[\frac{\cos 2(\mu-\nu)\tau_c}{(\mu-\nu)^2}+\frac{\cos
2(\mu+\nu+1)\tau_c}{(\mu+\nu+1)^2}],
$$
$$
S_n=\sum_{\nu=0}^\infty\frac{f_{2\nu+1}}{(2\nu+1)^4}\sin(2\nu+1)\tau_c,\qquad
\mu\ne \nu.
$$

Furthermore solution (5) is reduced to the superposition of sine
and cosine with extended modulated amplitudes. Note that a real
part of terms in braces is coefficient of cosine. In (5) function
$n(\tau)$ is absent. This technique as it is able  to derive the
frequency of prevalent oscillations $\nu_z\sim \lambda^2 N\gamma$
but amplitudes will be increased.

 The close equation was  also  examined in [6]. Keeping
in mind about the double integration of $n(\tau)$ let us introduce
by analogy to [6] new variables:
$$
p(t)=\int\!\!\int n d\tau d\tau, \quad x=\int\sqrt{p(\tau)}
d\tau,\quad v=\sqrt[4]{p(\tau)}\cdot z.
$$
After transformation Eq.(4) becomes
$$
\frac{d^2v}{dx^2}+\lambda^2
v=\frac{1}{16}\frac{v}{p^3}(4np-5p^{'2}).
$$
Here the left part has not a periodic coefficient but the solution
of inhomogeneous equation will be intricate.

Let us pass on to the Chasman-Green lattice. In this case there is
in centre focusing quadrupole of length $a_1$. Then  the cell 
contains, on both sides one after another
 straight sections with lengths $l_1$, $l$,$l_2$, between bending magnets of
length $d$, defocusing and 
 focusing quarupoles of length $a$.
 For one period $L$ is
$$
2d+4a+a_1+4l+2l_1+2l_2.
$$
 Corresponding equation for the vertical
oscillations can be expressed in the form
$$
\frac{d^2z}{d\tau^2}+\frac{C_1}{N^2}(\frac{a_1}{L}+\frac{2}{\pi}
\sum_{\nu=1}^{\infty}\frac{f_1}{\nu}\cos \nu\tau)z=0, \eqno (6)
$$
where $k=(L-2d)/2d,$
$$
f_1=4\sin \tau_2 a \sin\tau_2(a+l)\sin\tau_2(2a+l+2l_2)+
$$
$$
(-1)^{\nu}\sin\tau_2 a_1, \tau_2=\pi \nu/L.
$$
Contrary to the Eq.(4), given relation contains an additional
constant part along with the trigonometric series.

 An equation for triplet achromat cell 
is close in form to Eq.(6). Here there is in centre the defocusing
quadrupole of length $a_1$, then laterally through the free intervals are
 focusing quadrupole of length $a$  and bending magnet of length $d$. 
 For example, path of right side is equal to
$$
a_1+l_1+a+l_2+d+l_3,
$$
where $l_i$ is the length of straight sections.

In linear case the equation of axial oscillations becomes
$$
\frac{d^2z}{d\tau^2}+\frac{C_2}{N^2}[\frac{2a-a_1}{L}+\frac{2}{\pi}
\sum_{\nu=1}^{\infty}\frac{(-1)^{\nu}}{\nu}f_2\cos\nu \tau]z=0,
\eqno (7)
$$
where $C_2=\omega_q(1+k)/\omega_0$,
$$
 f_2=2\sin\tau_2 a\cos\tau_2(2l_1+a+a_1)-\sin\tau_2 a_1.
$$

Formulas (4)-(7) are unusual differential equations.
 In particulare, we cannot differentiate periodic coefficients,
as will be broken convergence of the series. Hill's method also
could not be used because the infinite determinant  increases for
$\lambda>>1$. This problem has a boundary layer, as 
quadrupoles operate in very narrow zones and are responsible for
the emergence of a small parameter at the highest derivative.
 But the derived equations
allow to perform simulation. Taking into account the
injection of particles and the state of the beam at specific points,
 we can  introduce the initial
conditions and to solve the Cauchy problem. Methods of this article 
can be used to study the motion of particles in more complex magnetic systems
including sextupoles and wigglers.

\newpage

REFERENCES

 [1] O.E. Shishanin, Journ.of Exper.and Teor.Phys. {\bf 90}, 725(2000).
 
 [2] O.E. Shishanin, Nucl.Inst.and Meth.in Phys.Res. A {\bf 558}, 74(2006). 

 [3] H. Wiedemann, Nucl.Inst.and Meth.in Phys.Res. A{\bf 246}, 4(1986). 

 [4] H. Wiedemann, Particle Accelerator Physics. Springer, 2007.

 [5] A.H. Nayfen, Introduction to Perturbation Techniques. John Wiley \& Sons, 1981.

 [6] A. Erdelyi, Asymptotic Expansions. Dover Books on Mathematics, 2010.

\end{document}